\documentclass[12pt]{iopart}

\usepackage{graphicx}
\usepackage{amsfonts}
\usepackage{amssymb}
\usepackage{amsbsy}
\usepackage{latexsym}
\def\p {\partial}

\def\be {\begin{equation}}
\def\ee  {\end{equation}}
\def\bea {\begin{eqnarray}}
\def\eea {\end{eqnarray}}
\def\nn {\nonumber}

\begin{document}

\title{ Solvable model for quantum gravity?}

\author{Jack Gegenberg and Viqar Husain}

\address{Department of Mathematics and Statistics\\University of New Brunswick\\
Fredericton, NB E3B 5A3, Canada}

\ead{geg@unb.ca, vhusain@unb.ca}

\begin{abstract}

We study a type of  geometric theory  with a  non-dynamical one-form field.  Its dynamical variables are an $su(2)$ gauge field and a triad of  $su(2)$ valued one-forms.  Hamiltonian decomposition reveals that the theory has  a true Hamiltonian, together with spatial diffeomorphism and Gauss law constraints, which generate the only local symmetries. Although perturbatively non-renormalizable, the model provides a test bed for the non-perturbative quantization techniques of loop quantum gravity.

\end{abstract}

% Keywords required only for MST, PB, PMB, PM, JOA, JOB?
%\vspace{2pc}
%\noindent{\it Keywords}: Article preparation, IOP journals
% Uncomment for Submitted to journal title message
%\submitto{\JPA}

 \section{Introduction}

It is well known that the perturbative approach to finding a theory of quantum gravity faces the problem of non-renormalizability.
This is partly solved in string theory,  but  at the expense of introducing extra dimensions. There exist however, non-renormalizable field theory models
for which the quantum theory is known to exist non-perturbatively.

One example of this is the Gross-Neveu model in three dimensions which is a
theory  with a four-fermion interaction; the interaction coupling constant  has negative mass dimension indicating power counting non-renormalizability. The model exists non-perturbatively in the ultraviolet regime \cite{3dgross-neveau-sol}. Another  is Einstein gravity in three dimensions \cite{witten3d,ashetal3d}. But this is a theory with no propagating degrees of freedom, and it is still not clear how to construct a consistent quantum theory of 3D gravity \cite{witten2}.  In any case, we do not know how to proceed from lower to higher dimensional theories, and so this case too may be considered special. Similar comments apply to BF theory \cite{vh-topqm,bg}.

 There is  so far only one example of a four dimensional diffeomorphism invariant theory
that is not renormalizable, but which exists as a non-perturbative quantum theory \cite{hk,ashetal-hk} . The catch is that although the model has local
degrees of freedom, its dynamics is trivial. It nevertheless shows that perturbative  non-renormalizability is not necessarily a sound criteria
for discarding a theory.

The question of  whether  this could be the case for quantum gravity in four dimensions  has been one motivation for  seeking a non-perturbative
canonical formulation.  However, no such program has yet been completed due to the problem of dealing with the Hamiltonian constraint in the
Dirac quantization approach, which leads to the Wheeler-DeWitt equation. Nevertheless there are recent indications that this can be circumvented
by deparametrizing the system by using matter fields to fix time and space gauges. One such approach uses a pressureless dust to fix only a
time gauge, thereby eliminating the Hamiltonian  constraint problem \cite{hp-dust} and replacing it by a true Hamiltonian with only spatial
diffeomorphism symmetry.

Motivated by such approaches we present a new type of theory with dynamical metric and a fixed one-form field\footnote{We note that there is a model with a  dynamical scalar field \cite{fb} to which the methods of this paper may be applied; in a particular time gauge, this model has an interesting  physical Hamiltonian with a  diffeomorphism constraint}.  The theory is such that  it has a built in time that does not arise via  a gauge fixing as in the aforementioned approaches. Its canonical decomposition reveals that there is a true Hamiltonian
together with spatial diffeomorphism and Gauss constraints, which generate the only gauge  symmetry.  The theory can be coupled to matter in natural way.
Its quantization can be carried out using the methods of loop quantum gravity. It therefore provides an example of a non-renormalizable geometric theory whose
quantum theory exists non-perturbatively.

In the following we describe the theory and its canonical formulation, and then outline a non-perturbative quantization scheme
using the background independent techniques developed in the loop quantum gravity (LQG) program.

\section{The model}

The fields in the theory are an $su(2)$ gauge field $A_\mu^i$, a dreibein $e_\mu^i$, scalar field $\phi$, and a fixed non-dynamical one-form
field $\zeta_\alpha$ which gives the two-form $\omega=d\zeta$. ($i,j,k\cdots$ are $su(2)$ indices, and $\alpha,\beta \cdots $ are
world indices.)  The dreibein fields $e_\mu^i$  define a degenerate $4-$metric, and give rise to the tensor density
\be
\tilde{u}^\alpha = \frac{1}{3!} \tilde{\eta}^{\alpha\beta\mu\nu}e_\beta^i e_\mu^j e_\nu^k\  \epsilon_{ijk},
\ee
where   $\tilde{\eta}^{\alpha\beta\mu\nu}$ is the Levi-Civita symbol (independent of $e_\alpha^i$ and $\zeta_\alpha$), and $\epsilon^{ijk}$  is the $su(2)$ structure constant. Using this
we define a scalar density  and vector field by
\be
\tilde{u} =  \tilde{u}^\alpha \zeta_\alpha, \ \ \ \ \ \  u^\alpha = \frac{\tilde{u}^\alpha} {\tilde{u}}, \label{u}
\ee
and a  co-triad by
\be
e^{\alpha}_i = \frac{1}{2\tilde{u}} \ \tilde{\eta}^{\alpha\beta\mu\nu}\zeta_\beta e_\mu^j e_\nu^k\  \epsilon_{ijk}.  \label{cotriad}
\ee
The scalar density $\tilde{u}$ would vanish if $\zeta_\alpha$  were a linear combination of the $e_\alpha^i$, so we assume
this is not the case. These definitions give the relations \footnote{With these relations we note that the 2-form $\omega$ is invertible  (because $u\cdot \omega ={\cal L}_u \zeta \ne 0$, and $e^{\alpha}\cdot \omega \ne 0$), therefore it is a symplectic form. However this fact is not needed in our subsequent development of the model.}
\be
u^\alpha \zeta_\alpha = 1,\ \ \ \ \ \  u^\alpha e_\alpha^i = 0, \ \ \ \ \  \zeta_\alpha e^{\alpha}_i=0
\ee
\be
e^{\alpha}_i e_\alpha^j = \delta_i^j, \ \ \ \ \  e^{\alpha}_i e_\beta^i = \delta^\alpha_\beta.
\ee
We note finally that a non-degenerate Euclidean or Lorentzian signature $4-$metric may be defined by
\be
g_{\alpha\beta} = \pm \zeta_\alpha \zeta_\beta + e_\alpha^i e_\beta^i.
\label{4metric}
\ee
We are now ready to define the action for the model which contains the field $\zeta_\alpha$ as a fixed ``background'' structure. The action is
\bea
 S&=&  S_G + S_\Lambda + S_\phi \nn\\
 &=& \frac{1}{l^2}\int_M  \tilde{\eta}^{\alpha\beta\mu\nu} \epsilon^{ijk}e_\alpha^ie_\beta^j  F_{\mu\nu}^k(A)  + \Lambda \int_M \tilde{u} \\
&& +  \int_M \tilde{u}\ \left( -u^\mu u^\nu\p_\mu\phi \p_\nu\phi + e^{\mu}_ie^{\nu}_i \p_\mu\phi \p_\nu \phi\right).
\eea
The first term is the action of the model introduced in \cite{hk}, where $F(A)$ is the curvature of the gauge field $A$.  Its canonical theory has an identically vanishing Hamiltonian  constraint, so it is a theory with three local degrees of freedom and no dynamics. The fixed one-form field $\zeta_\alpha$  makes it possible to introduce the cosmological constant term
and coupling to matter in the manner displayed.   The coupling constant $l$ is a fundamental length scale obtained by assigning the  usual canonical
dimension to the connection, i.e. $A$ has mass dimension one, and $e$  is dimensionless. This assignment makes the theory power counting non-renormlizable
just as in Einstein gravity, since changing the gauge algebra from $so(3,1)$ to $su(2)$ does not affect this counting.

\subsection{Hamiltonian theory}
To construct the Hamiltonian theory let  us introduce the embedding variable  $X^\alpha (t,x^a)$ which provides a smooth map
\be
X:   \mathbb{R}\times \Sigma \longrightarrow M
\ee
where $\Sigma$ is a three manifold. The inverse map gives the functions $x^a(X)$ and $t(X)$.  The 3+1 split of the first term in the action
is obtained \cite{hk} by substituting into the action the decompositions
\be
\tilde{\eta}^{\alpha\beta\mu\nu} = \tilde{\eta}^{abc}\dot{X}^\alpha X^\beta_{,a}X^\mu_{,b}X^\nu_{,c},
\ee
where the time deformation vector field
$\dot{X}^\alpha$ decomposes as
\be
\dot{X}^\alpha = u^\alpha + N^\alpha = u^\alpha + X^\alpha_{,a}N^a. \label{Xdot}
\ee
We also use the  spatial projections of the fields defined by
\be
e_a^i = e_\alpha^i X^\alpha_{,a}, \ \ \  A_a^i = A_\alpha^i X^\alpha_{,a}, \ \ \  \tilde{e}^{ai} =  \tilde{\eta}^{abc}e_b^je_c^k\epsilon^{ijk}, \ \ \ e^{ai}=\tilde{e}^{ai}/\tilde{e},
\ee
where $\tilde{e}= \tilde{\eta}^{abc}e_a^ie_b^je_c^k\epsilon^{ijk}$.
These are the decompositions needed to arrive at the canonical form of the first part of the action, which is
\be
S_G = \int_M d^3x dt\left[\tilde{e}^{ai} \dot{A}_a^i   - N^a(\p_{[a}A_{b]}^i\tilde{e}^{bi} -A_a^i\p_b \tilde{e}^{bi}) - \Lambda^i D_a\tilde{e}^{ai} \right]
\ee
where $N^a = e^{ai}(e_\beta^i \dot{X}^\beta)$ and $\Lambda^i = A_\alpha^iu^\alpha$. This identifies the fundamental Poisson brackets for the geometric variables:
\be
\{A^i_a(x),\tilde{e}^b_j(x')\}=\tilde{\delta}^3(x-x')\delta^i_j
 \delta^b_a.
\ee

To obtain the canonical  decomposition of $S_\Lambda$ and $S_\phi$ we note first that
 \bea
\tilde{u} &=& \dot{X}^\alpha \zeta_\alpha \tilde{e} = (1 + X^\alpha_{,a}\zeta_\alpha N^a)\tilde{e} , \nn\\
e^{\alpha i} &=&  X^\alpha_{,a}e^{ai} + \dot{X}^\alpha (t_{,\beta}e^{\beta i}).
\eea
Now the identity $e^{\alpha i} \zeta_\alpha =0$ applied to the last equation gives
\be
0=  X^\alpha_{,a} \zeta_\alpha  e^{ai} + \dot{X}^\alpha \zeta_\alpha (t_{,\beta}e^{\beta i}).
\ee
Thus if we choose the foliation $X^\alpha(t,x^a)$ such that  $X^\alpha_{,a} \zeta_\alpha=0$ (ie. adapted to the
fixed field $\zeta_\alpha$) we have
\be
\tilde{u} = \tilde{e}, \ \ \ \ \ \ \  e^{\alpha i} = X^\alpha_{,a}e^{ai}.
\ee
Substituting these together with (\ref{Xdot}) into the action gives
\be
S_\Lambda + S_\phi = \int d^3xdt\left[ \Lambda \tilde{e} + \frac{P^2_\phi}{2\tilde{e}} + \tilde{e}e^{ai}e^{bi}\p_a\phi \p_b\phi  -N^a P_\phi \p_a\phi \right].
\ee

The Hamiltonian decomposition of the full action then shows that the phase space variables are the canonical pairs
$(e^{ai}, A_a^i)$ and $(\phi, P_\phi)$ with the Hamiltonian
\be
H = \int d^3x \left[ \Lambda \tilde{e} + \frac{P^2_\phi}{4\tilde{e}} + \tilde{e}e^{ai}e^{bi}\p_a\phi \p_b\phi\right].\label{cham}
\ee
 The theory has two sets of first class constraints that generate SU(2) gauge transformations and  spatial diffeomorphisms. Thus the theory
 has four local configuration degrees of freedom of which three are geometric and one is matter. The remarkable feature is that true dynamics is obtained
 by introducing a fixed one-form field which may be interpreted as providing a symplectic structure on the manifold. Thus the presence of this structure
 may be viewed as providing a time variable, while maintaining full general covariance of the action.

 The Hamiltonian equations of motion provide a view of the dynamics. Evolution is a combination of gauge (Gauss and spatial diffeomorphisms) and true motion via $H$.  We note first that the three geometry does not evolve:
 \be
 \dot{\tilde e}^{ai} = \{ \tilde{e}^{ai},  H\} = 0,
 \ee
 but its conjugate connection does
 \be
 \dot{A}_a^i = \{ A_a^i, H \} = e_a^i\left(  \Lambda- \frac{P_\phi^2}{4 \tilde{e}^2} -   e^{bj}e^{cj}\p_b\phi \p_c\phi \right)+2\p_a\phi e^b_i\p_b \phi.
 \ee
 The scalar field equations are the usual ones for a field on a curved space-time given by the metric (\ref{4metric}).

The geometrical phase space variables in our theory are identical to those of the Ashtekar-Barbero canonical formulation of general relativity. There the connection $A_a^i$ is a sum of the (spatial) metric connection and the  extrinsic curvature. Thus the comparison allows us to interpret the canonical equations of motion of our model as evolving the extrinsic curvature, but not the spatial metric. With this in mind, the model may be viewed as evolving both the matter field and the four-geometry (through the connection $A_a^i$).

\section{Quantization}

Using the similarity of the geometrical part of the phase space with that of general relativity in the connection-triad variables, we turn to a discussion
of the quantum theory of this model. Using an extension of the spin network Hilbert space  used in loop quantum gravity to include scalar matter degrees
of freedom \cite{thiemann-qsd5}, we will see that it is possible to set up a complete quantum theory.

The starting point of the LQG approach is the set of phase space functions
\be
 U_\gamma[A] = P \exp\int_\gamma A_a^i\tau^i dx^a,\ \ \ \ \ \ \  F^i_S = \int_S  \tilde{e}_a^i dS^a,
\ee
where $\gamma$ is a loop and $S$  a surface in a spatial hypersurface $\Sigma$, and $\tau^i$ are generators of $SU(2)$. Gauge invariant versions of these were first used for quantization of BF theory \cite{vh-topqm} and in \cite{bg}. Their Poisson bracket forms the so called holonomy-flux algebra
\be
\{ H_\gamma[A], F^i_S   \} =  \int_\gamma ds \int_S\  d^2\sigma U_\gamma[A]\tau^i \delta^3(\gamma(s), S(\sigma)).
\ee
The analogous observables for the scalar field are
\be
 V_k(\phi(x)) = \exp [ik\phi(x)], \ \ \ \ \ \ \ \   P_f = \int_\Sigma f P_\phi d^3x,
\ee
where $k \in \mathbb{R}$ and $f(x)$ is a suitable function with rapid fall-off.  These satisfy
\be
\{ V_k (x), P_f \} = ik f(x)V_k(x).
\ee

\subsection{Geometry Hilbert space}
There is a well-defined path to quantization of the gravitational variables which are discussed in detail in a number of reviews \cite{lqg-revs}. Therefore we restrict attention to describing the basic guidelines.  A crucial first step is the choice of Hilbert space for a connection representation $\Psi[A]$. One considers an oriented graph $\Gamma$ with  ordered edges $e_1, e_2 \cdots e_N$, and vertices $n_1,n_2\cdots n_M$ embedded in the spatial surface $\Sigma$, and associates the holonomy function in the representation $j$ of SU(2), $H_{e}^j[A]$,  with edge $e$.    A spin network  state is a function of such holonomies
\be
f[A] = f(U_{e_1}^{j_1}, U_{e_2}^{j_2},\cdots  U_{e_N}^{j_N}).
\ee
These are essentially functions of $SU(2)$ group elements, so the natural inner product is  the Haar measure on (tensor product  copies) of this group. A convenient orthonormal basis for this space of functions is the spin network basis; the wave function of a graph with a single edge $e$ is the matrix  $(U_e^j)[A]_{m_1,m_2} \equiv \langle A| j;m_1,m_2\rangle$
in the representation $j$, where $m_1,m_2$ are its matrix indices. This generalizes readily to multi-edges graphs.

The space ${\cal H}_{Kin}$ is not the physical Hilbert space of our model, since its elements are neither gauge nor spatial diffeomorphism invariant. The LQG path to achieve invariance under these transformations is done in two steps. The first step is the formulation of ${\cal H}_{Kin}^G$, the space of SU(2) invariant states. The usual formulation of this involves group averaging of  states  in   ${\cal H}_{Kin}$. Intuitively this amounts to tracing over matrix indices using SU(2) invariant tensors (called intertwiners) at all vertices of the graph $\Gamma$, and ensuring there are no open ``dangling'' edges. This gives the Gauss invariant states.  Such states may be represented as the kets
\be
|\Gamma;{\bf j};{\bf I}\rangle:=|\Gamma;  j_1,  \cdots j_N; I_1,  \cdots I_M\rangle,
\ee
where a spin $j$ is associated with each edge and an intertwiner $I$ with each vertex of the graph $\Gamma$. The inner product in ${\cal H}_{Kin}^G$ is the obvious one guided by this characterization of the basis:
\bea
 &&\langle \Gamma; j_1, \cdots j_N; I_1,  \cdots I_M   |\Gamma;  j_1',  \cdots j_N'; I_1', \cdots I_M'\rangle \nn\\
 &&= \delta_{j_1,j_1'}\cdots \delta_{j_N,j_N'}
 \delta_{I_1,I_1'}\cdots \delta_{I_M, I_M'},
\eea
if the graphs are the same, and zero otherwise. For spin networks with only trivalent vertices, the intertwiners are unique (up to a multiplicative constant). An explicit example of a gauge invariant trivalent spin network state with three edges is
\be
\psi[A]_{1,\frac{1}{2},\frac{1}{2}} =  \langle A| 1, {\textstyle\frac{1}{2}}, {\textstyle\frac{1}{2}}; \sigma, \sigma\rangle=[U^{\frac{1}{2}}_{e_1}]^A_{\ B} \  [U^{1}_{e_2}]^i_{\ j} \ [U^{\frac{1}{2}}_{e_3}]^C_{\ D}\  \sigma_{iAC}\  \sigma^{jBD},
\label{gspin}
\ee
where $\sigma^i$ are the Pauli matrices. This example also illustrates why edges must be oriented; the matrix indices $(iAC)$ come together at one vertex and $(jBD)$ at the other.

Having characterized ${\cal H}_{Kin}^G$ in this manner, the next step is to address the  requirement of invariance under spatial diffeomorphisms.  We note first that there is a natural action of diffeomorphisms on the gauge invariant spin network states such as (\ref{gspin}). This stems from the observation that such transformations ``drag the graph around'' but do not affect the combinatoric information in the spins and intertwiners \cite{ashetal-hk,lqg-revs}. Formally, for $\phi \in { Diff}(\Sigma)$, we have
\be
{\cal {U}}_D[\phi] |\Gamma;  j_1,  \cdots j_N; I_1,  \cdots I_M\rangle = |\phi^{-1}\Gamma;  j_1,  \cdots j_N; I_1,  \cdots I_M\rangle.
\ee
Thus for a fixed graph $\Gamma$ the diffeomorphism invariant information  is just the set of spins and intertwiners (up to some subtleties \cite{lqg-revs}). We denote this Hilbert space by ${\cal H}_{geom}$, and in the following consider the case where the underlying graph is a cubic (abstract) lattice. Thus each node will be 6-valent, and we will assume that the associated non-zero spins and intertwiners form a  finite set. This will aid in defining the physical Hamiltonian operator.\footnote{The choice of cubic graph represents a restriction of the quantum theory, since in principle all graphs should be included; this choice allows a systematic construction of the Hamiltonian operator. The solution of the diffeomorphism constraint to yield ${\cal H}_{geom}$ for a cubic lattice proceeds as in \cite{ashetal-hk}, with a finite set of excitations on the lattice.}

\subsection{Matter Hilbert space}

The geometry Hilbert space ${\cal H}_{geom}$ described above is the physical Hilbert space of the model without matter. Its extension to include matter is accomplished by associating an additional quantum number with the vertices of graphs. Given a graph $\Gamma$ with vertices $v_1\cdots v_M$, a basis  for the matter Hilbert space, (${\cal H}_{matter}$) is $|k_1, \cdots k_M\rangle$, where $k_i \in \mathbb{R}$ are the quantum numbers associated with matter. The inner product is
\be
\langle k'_1, \cdots,k'_M |k_1, \cdots,k_M\rangle=\delta_{k'_1,k_1}\cdots \delta_{k'_M,k_M}.
\ee
The classical scalar field variables $V_k(\phi(x)),P_f$ defined above have the quantum realizations
\bea
\hat{V}_{k}(v_l)|k_1,\cdots,k_M\rangle &=&|k_1,\cdots,k_l+k ,\cdots,k_M\rangle,\\
\hat{P}_f|k_1,\cdots,k_M\rangle &=& \sum_{i=1}^M k_i f(v_i)|k_1,\cdots,k_M\rangle,
\eea
where $v_i$ is a vertex.  It is readily verified that these definitions provide a representation of the classical Poisson algebra.

\subsection{Physical Hilbert space and Hamiltonian}

The physical Hilbert space of our model is the tensor product ${\cal H}_{geom} \otimes {\cal H}_{matter}$, with basis
\be
|\Gamma;{\bf j};{\bf I};{\bf k}\rangle
=|\Gamma;  j_1,  \cdots, j_N; I_1,  \cdots, I_M; k_1, \cdots,k_M\rangle.
\ee
As mentioned above we assume that the geometric and matter excitations are on an infinite cubic graph. Its regularity provides a systematic way to construct  the Hamiltonian operator to which we now turn.

The classical expression for the Hamiltonian  (\ref{cham}) contains geometric terms that appear in the Hamiltonian constraint of LQG. The operator realizations of these  are well studied in the literature \cite{thiemann-qsd5}. For example the $\tilde{e}$ term in the Hamiltonian is realized using the LQG volume operator, and its inverse is realized as a commutator of the square root of the volume and holonomy operators, a construction well known in LQG.

Turning to the matter operators, the  $P_\phi^2$ factor is diagonal in the basis we are using. It can be localized by writing the integral for $P_f$ as a sum over vertices of the graph, taking $f$ to be unity, ie
\be
\int d^3x \frac{P^2_\phi}{\tilde{e}}  \longrightarrow \sum_{i} {\widehat{\frac{1}{\tilde{e}}}}_{v_i} P^2(v_i)
\ee
The factors of $\partial_a\phi$ may be realized by using a ``finite difference'' approach. We first define a local field operator as
\be
\Phi_k(v_i) := \frac{1}{2ik}\  \left( V_k(v_i) -  V_{-k}(v_i)\right).
\ee
Using this, one way to define the operator corresponding to the matter gradient $e^a\p_a\phi$  via a finite difference scheme. The simplest such scheme is forward Euler, where for a single direction $z_k$ on the cubic lattice we have
\be
e^z\p_z \phi(v_i) \longrightarrow \hat{F}_{z_k}(\Phi_k(v_{i+z_k}) -  \Phi_k(v_k)),
\ee
where $\hat{F}_{z_k}$  is the flux operator associated with the edge $z_k$ that connects the adjacent vertices $v_{i+z_k}$ and $v_k$.  It is evident that there are other ways to write this operator; our purpose is to point out that the Hamiltonian can be defined using the basic operators.

\section{Summary}
We have developed a new type of geometric theory defined on a symplectic manifold that is topologically $\mathbb{R}^4$. The theory has a ``built in'' time that does not arise via  a gauge fixing as in the aforementioned approaches. Its canonical decomposition reveals that there is a true Hamiltonian together with spatial diffeomorphism and Gauss constraints, which generate the only gauge  symmetry.  The theory can be coupled to matter in a natural way. The connection $A_a^i$  defines an extrinsic curvature via the Ashtekar-Barbero relation $A_a^i = \Gamma_a^i(e) + K_a^i$. From this we note the theory may be interpreted as giving a dynamical 4-geometry, even though the 3-geometry given by $e^a_i$ does not evolve.  Quantization of the theory can be carried out using the methods of LQG. The model therefore provides an example of a perturbatively non-renormalizable geometric theory that exists non-perturbatively at the quantum level.

 \section{Acknowledgements}
 This work was supported by the Natural Science and Engineering Research  Council of Canada.

\end{document}